


\hsize=5.5in \hoffset=0.5in
\vsize=220truemm \voffset=2truemm

\font\tenssi=cmssi10
\def\ssi{\tenssi}

\hfuzz=0.5em\vfuzz=1ex
\overfullrule=0em
\tolerance=300


\def\+{\eqprefix\number\eqncnt}  
\def\e{\global\advance\eqncnt by 1 \+}    
\def\en#1{\e\xdef#1{\+}}   
\def\eqprefix{\rm}                
\newcount\eqncnt \eqncnt=0

\def\ref#1#2{\unskip 
            {\def\temp{$^{\hbox{\sevenrm #1}}$}%
                            \def\
{ }\def\ { }
                            \if #2..\temp\spacefactor3000 \gdef\reftemp{}\else
                            \if #2,,\temp\spacefactor1250 \gdef\reftemp{}\else
                              \temp\gdef\reftemp{#2}\fi\fi}\reftemp}
\def\Ref. #1.{\par\vfil\penalty -100\vfilneg
                    \noindent\hangindent 20pt $^{#1}$\ignorespaces}
\newcount\refcnt\refcnt=0
\def\refdef#1{\global\advance\refcnt by 1
              \xdef#1{\number\refcnt}}
\refdef\landau
\refdef\beliaev
\refdef\rosenbluth
\refdef\trubnikov
\refdef\lifshitz
\refdef\klimontovich
\refdef\bezzerides
\refdef\franz


\def\pmb#1{{\setbox0=\hbox{#1}%
  \kern-.03em\copy0\kern-\wd0\raise.06em\copy0\kern-\wd0
  \kern .06em\copy0\kern-\wd0\raise.06em\box0 }}

\def\bt#1{\pmb{\ssi#1\/}} 

\def\frac#1#2{{#1\over#2}}

\def\d{\partial}

\def\pd#1#2{\mathchoice{\d#1\over\d#2}{\d#1/\d#2}{\d#1/\d#2}{\d#1/\d#2}}

\def\wg{\varphi}
\def\wh{\psi}

\def\DD{{\bt D}}
\def\DC{{\cal D}}
\def\F{{\bf F}}
\def\GC{{\cal G}}
\def\HC{{\cal H}}
\def\II{{\bt I}}
\def\K{{\bf K}}
\def\KC{{\cal K}}
\def\LL{{\bt L}}
\def\LC{{\cal L}}
\def\s{{\bf s}}
\def\u{{\bf u}}
\def\UU{{\bt U}}
\def\v{{\bf v}}

\def\jour#1#2(#3)#4*%
  {{\frenchspacing#1\unskip}~{\bf#2\unskip}, #4 (#3)}


\rightline{PPPL--2467 (Aug.~1987)}
\rightline{Phys.\ Rev.\ Lett.\ {\bf 59}(16), 1817--1820 (Oct.~1987)}
\bigskip
\centerline{DIFFERENTIAL FORM OF THE COLLISION INTEGRAL}
\centerline{FOR A RELATIVISTIC PLASMA}
\bigskip
\centerline{Bastiaan J. Braams and Charles F. F. Karney}
\medskip
\centerline{%
Plasma Physics Laboratory, Princeton University, Princeton, NJ~08544.}

\bigskip

{\sl\noindent
The differential formulation of the Landau-Fokker-Planck collision
integral is developed for the case of relativistic electromagnetic
interactions.\par}
\smallskip
{\sl\noindent PACS numbers 52.25.Dg, 52.20.$-$j, 52.60.$+$h}

\medskip

Kinetic theory is founded upon the Boltzmann equation, which is a conservation
equation for the phase-space distribution function of each species in an
ensemble of interacting particles. For the case of Coulomb interactions, Landau
\ref\landau\ expressed the collision term in the Fokker-Planck form.
This mixed integro-differential representation was extended to relativistic
electromagnetic interactions by Beliaev and Budker \ref\beliaev. For the
nonrelativistic case, it was shown by Rosenbluth et~al. \ref\rosenbluth\ and by
Trubnikov \ref\trubnikov\ that the integrals appearing in the collision term
can be expressed in terms of the solution of a pair of differential equations.
The present work extends that formulation to the relativistic collision
integral. Using an expansion in spherical harmonics the relativistic
differential formulation is then applied to calculate the scattering and
slowing down of fast particles in a relativistic equilibrium background plasma.
Our work is relevant to the study of high temperature plasma in fusion energy
research and in astrophysics.

In the work of Landau \ref\landau\ and that of Beliaev and Budker
\ref{\beliaev,\lifshitz}, the collision term that occurs on the right-hand
side of the Boltzmann equation for species~$a$ and describes the effect of
collisions with species~$b$ is written in the Fokker-Planck form,
$$C_{ab}=\pd{}{\u}\cdot(\DD_{ab}\cdot\pd{f_a}{\u}-\F_{ab}f_a),\eqno(\e)$$
in which the coefficients $\DD_{ab}$ and $\F_{ab}$ are defined by
$$\eqalignno{
  \DD_{ab}(\u)&=\frac{q_a^2q_b^2}{8\pi\epsilon_0^2m_a^2}\log\Lambda_{ab}
  \int\UU(\u,\u')f_b(\u')\,d^3\u',&(\en\dfeq a)\cr
  \F_{ab}(\u)&=-\frac{q_a^2q_b^2}{8\pi\epsilon_0^2m_am_b}\log\Lambda_{ab}\int
  \biggl(\pd{}{\u'}\cdot\UU(\u,\u')\biggr)f_b(\u')\,d^3\u'.&(\+b)\cr}$$
Here, $f_a$ and $f_b$ are the distribution functions for the two species, $\u$
is the ratio of momentum to species mass, $q_a$ and $q_b$ are the species
charge, $m_a$ and $m_b$ are the species mass, $\epsilon_0$ is the vacuum
dielectric permittivity, and $\log\Lambda_{ab}$ is the Coulomb logarithm. The
kernel $\UU$ is specified below. This form of the collision operator is only
approximate because of the introduction of cutoffs in the collision integral.
More accurate operators that take into account Debye shielding at large impact
parameters and large-angle scattering and quantum effects at small impact
parameters have been derived \ref{\klimontovich,\bezzerides}. The purpose of
this letter is to present a differential formulation for the integral
transforms that occur in Eqs.~(\dfeq). To avoid unnecessary clutter we discard
the factor that depends only on the species properties, drop the species
subscript, and consider the transforms
$$\eqalignno{
  \DD(\u)&=\frac1{8\pi}\int\UU(\u,\u')f(\u')\,d^3\u',&(\en\dfnorm a)\cr
  \F(\u)&=-\frac1{8\pi}\int\biggl(\pd{}{\u'}\cdot\UU(\u,\u')\biggr)
    f(\u')\,d^3\u'.&(\+b)\cr}$$

For guidance, let us recall briefly the nonrelativistic theory
\ref{\rosenbluth,\trubnikov}. In that case the momentum-to-mass ratios $\u$
and $\u'$ reduce to the velocities $\v$ and $\v'$, and the collision kernel is
the one given by Landau \ref\landau,
$\UU=(|\s|^2\II-\s\s)/|\s|^3$, where $\s=\v-\v'$.
It may be seen that $\UU=\d^2|\s|/\d\v\d\v$ and
$(\d/\d\v')\cdot\UU=-2\d|\s|^{-1}/\d\v$.
To obtain the differential formulation, these representations are inserted
into Eqs.~(\dfnorm), and the differentiation with respect to
$\v$ is moved outside the integration over $\v'$. Defining the potentials
$h(\v)=-(1/8\pi)\int|\s|f\,d^3\v'$ and
$g(\v)=-(1/4\pi)\int|\s|^{-1}f\,d^3\v'$, we have
$\DD=-\d^2h/\d\v\d\v$ and $\F=-\d g/\d\v$. Furthermore, from
$\Delta|\s|=2|\s|^{-1}$ and $\Delta|\s|^{-1}=-4\pi\delta(\s)$
it follows that $h$ and $g$ obey the equations $\Delta h=g$ and $\Delta g=f$.
($\Delta$ denotes the Laplacian with respect to the variable $\v$.) These
equations provide the differential formulation of the collision term in the
nonrelativistic case.

The Landau collision kernel was obtained in a semi-relativistic fashion,
assuming Coulomb collisions and relativistic particle kinematics. It is a good
approximation to the fully relativistic kernel given below provided that
$|\v.\v'|\ll c^2$, which is true when one of the colliding particles is
nonrelativistic. However, the reduction of the collision integral to the
differential form of Rosenbluth and Trubnikov relies on the stronger
assumptions $|\v|^2\ll c^2$ and $|\v'|^2\ll c^2$, and is therefore entirely
nonrelativistic. A differential formulation that is exactly equivalent to
the Landau collision integral was given by Franz \ref\franz.

We turn now to the differential formulation of the relativistic collision
integral due to Beliaev and Budker \ref{\beliaev,\lifshitz,\klimontovich}.
They obtained the expression
$$\UU(\u,\u')=\frac{r^2/(\gamma\gamma')}{(r^2-1)^{3/2}}
  \bigl((r^2-1)\II-\u\u-\u'\u'+r(\u\u'+\u'\u)\bigr),\eqno(\e a)$$
in which $\gamma=\sqrt{1+|\u|^2}$, $\gamma'=\sqrt{1+|\u'|^2}$, and
$r=\gamma\gamma'-\u.\u'$. (We set $c=1$ in this part of the paper.) One finds
$$\pd{}{\u'}\cdot\UU(\u,\u')
  =\frac{2r^2/(\gamma\gamma')}{(r^2-1)^{3/2}}(r\u-\u').\eqno(\+b)$$
Notice that $r$ is the relativistic correction factor for the relative velocity
between the two particles (i.e., for the velocity of one particle in the rest
frame of the other). Conversely, this relative velocity is given by
$r^{-1}(r^2-1)^{1/2}$.

In developing a differential formulation for the collision term based on the
Beliaev and Budker kernel, it is helpful to work in terms of relativistically
covariant quantities. The expression $\gamma\gamma'\UU$ is equal to the space
part of a four-tensor $W$ that depends on the four-vectors $u=(\gamma,\u)$ and
$u'=(\gamma',\u')$,
$$W^{ij}(u,u')=\frac{r^2}{(r^2-1)^{3/2}}\bigl((r^2-1)g^{ij}
  -u^iu^j-u'^iu'^j+r(u^iu'^j+u'^iu^j)\bigr),\eqno(\e a)$$
where $g^{ij}$ is the metric tensor, with signature ${-}{+}{+}{+}$.
($r=-u_iu'^i$ is clearly a four-scalar.) The tensor $W$ is symmetric
($W^{ij}=W^{ji}$), symmetric in $u$ and $u'$, satisfies $u_iW^{ij}=0$, and
satisfies $W^i_i=2r^2(r^2-1)^{-1/2}$. Likewise
$\gamma\gamma'(\pd{}{\u'})\cdot\UU$ is the space part of the four-vector $V$,
where
$$V^i(u,u')=\frac{2r^2}{(r^2-1)^{3/2}}(ru^i-u'^i).\eqno(\+b)$$

If the relativistic differential formulation is to parallel most closely the
nonrelativistic formulation, then one should find a representation of the form
$W^{ij}=\HC^{ij}\wh$ and $V^i=-2\GC^i\wg$, where $\wh$ and $\wg$ are
four-scalars depending on $u$ and $u'$, and $\HC^{ij}$ and $\GC^i$ are
covariant differential operators acting on the variable $u$. In the
nonrelativistic limit, $\wh$ should reduce to $|\v-\v'|$ and $\wg$ should
reduce to $|\v-\v'|^{-1}$. It should be possible to transform $\wh$ and $\wg$
to delta functions by a sequence of second-order differential operators. The
potentials would be defined as $h=-(1/8\pi)\int(\wh f/\gamma')\,d^3\u'$ and
$g=-(1/4\pi)\int(\wg f/\gamma')\,d^3\u'$; these expressions define
four-scalars (cf.\ Ref.~\lifshitz). The differential equations satisfied by
$h$ and $g$ follow immediately from those satisfied by $\wh$ and $\wg$.
Finally, $\DD$ would be obtained as the space part of $-\gamma^{-1}\HC^{ij}h$
and $\F$ as the space part of $-\gamma^{-1}\GC^ig$. In fact, it will turn out
that the relativistic formulation has to be somewhat more complicated, but not
fundamentally different from the outline just sketched.

A function of the four-vectors $u$ and $u'$ that is a four-scalar must be a
function of $r=-u.u'$ alone. The form of the differential operators $\HC^{ij}$
and $\GC^i$ is restricted because these should be interior operators on the
surface $u^2=-1$ in four-space. In addition, it is required that
$\HC^{ij}=\HC^{ji}$ and $u_i\HC^{ij}=0$. Under those restrictions it is
found that the most general form of $\HC^{ij}$ and $\GC^i$, up to a
multiplicative constant, is
$\HC^{ij}\chi=\LC^{ij}\chi+\alpha(g^{ij}+u^iu^j)\chi$
and $\GC^i\chi=\KC^i\chi+\beta u^i\chi$.
Here, $\alpha$ and $\beta$ are arbitrary constants, and
$$\eqalignno{
  \LC^{ij}\chi&=(g^{ik}+u^iu^k)(g^{jl}+u^ju^l)\frac{\d^2\chi}{\d u^k\d u^l}
    +(g^{ij}+u^iu^j)u^m\pd{\chi}{u^m},&(\e a)\cr
  \KC^i\chi&=(g^{ik}+u^iu^k)\pd{\chi}{u^k}.&(\+b)\cr}$$
The spatial part of $\LC^{ij}\chi$ is $\LL\chi$ and that of $\KC^i\chi$ is
$\K\chi$ where
$$\eqalignno{
  \LL\chi&=\gamma^{-2}\frac{\d^2\chi}{\d\v\d\v}-\v\pd\chi\v-\pd\chi\v\v,
    &(\en\opdef a)\cr
  \K\chi&=\gamma^{-1}\pd\chi{\v},&(\+b)\cr}$$
in which $\v=\u/\gamma$, and $\d/\d\v=\gamma(\II+\u\u)\cdot\d/\d\u$.
If $\chi$ is a function of $r$ alone then
$$\LC^{ij}\chi=\frac{d^2\chi}{dr^2}(ru^i-u'^i)(ru^j-u'^j)
  +r\frac{d\chi}{dr}(g^{ij}+u^iu^j)$$
and $\KC^i\chi=(d\chi/dr)(ru^i-u'^i)$.
One is thereby led to the representations
$$\eqalignno{
  W^{ij}&=\bigl[\LC^{ij}+g^{ij}+u^iu^j\bigr]\sqrt{r^2-1}\cr
  &\qquad{}-\bigl[\LC^{ij}-g^{ij}-u^iu^j\bigr]
      \bigl(r\cosh^{-1}\!r-\sqrt{r^2-1}\bigr),&(\e a)\cr
  V^i&=-2\KC^i\bigl(r(r^2-1)^{-1/2}-\cosh^{-1}\!r\bigr).&(\+b)\cr
}$$

These representations for $W$ and $V$ are only suitable for constructing
a differential formulation of the collision term if the functions that occur
on the right-hand sides can be reduced to delta functions by some sequence
of differential operators. For that purpose the contraction $L=\LC^i_i$
is needed; in terms of the three-space variables it is
$$L\chi=(\II+\u\u):\frac{\d^2\chi}{\d\u\d\u}+3\u\cdot\pd\chi\u.
\eqno(\en\laplace)$$
If $\chi$ is a function of $r$ alone, then
$L\chi=(r^2-1)(d^2\chi/dr^2)+3r(d\chi/dr)$ away from $r=1$; at $r=1$
(or $\u=\u'$) there
may be a singularity. Specifically, it is found that
$$\eqalign{
  L\bigl(r(r^2-1)^{-1/2}\bigr)
    &=-4\pi\gamma\delta(\u-\u'),\cr
  [L+1]\bigl((r^2-1)^{-1/2}\bigr)
    &=-4\pi\gamma\delta(\u-\u'),\cr
  L\bigl(\cosh^{-1}\!r\bigr)
    &=2r(r^2-1)^{-1/2},\cr
  [L-3]\bigl(\sqrt{r^2-1}\bigr)
    &=2(r^2-1)^{-1/2},\cr
  [L-3]\bigl(r\cosh^{-1}\!r-\sqrt{r^2-1}\bigr)
    &=4\sqrt{r^2-1}.\cr
}$$

The explicit form of the differential representation of Eqs.~(\dfnorm) based on
the Beliaev and Budker collision kernel \ref\beliaev\ follows: The potentials
are
$$\eqalignno{
  h_0&=-(1/4\pi)\int(r^2-1)^{-1/2}f(\u')/\gamma'\,d^3\u',&(\en\pot a)\cr
  h_1&=-(1/8\pi)\int\sqrt{r^2-1}\,f(\u')/\gamma'\,d^3\u',&(\+b)\cr
  h_2&=-(1/32\pi)\int\bigl(r\cosh^{-1}\! r-\sqrt{r^2-1}\bigr)
    f(\u')/\gamma'\,d^3\u',&(\+c)\cr
  g_0&=-(1/4\pi)\int r(r^2-1)^{-1/2}f(\u')/\gamma'\,d^3\u',&(\+d)\cr
  g_1&=-(1/8\pi)\int\cosh^{-1}\! r\,f(\u')/\gamma'\,d^3\u'.&(\+e)\cr
}$$
These potentials satisfy the differential equations
$$\eqalign{
[L+1]h_0&=f,\cr [L-3]h_1&=h_0,\cr [L-3]h_2&=h_1,\cr}\qquad
\eqalign{Lg_0&=f,\cr Lg_1&=g_0.\cr}\eqno(\en\diffeqs)$$
Finally one obtains $\DD$ and $\F$ as
$$\eqalignno{\DD(\u)&=-\gamma^{-1}[\LL+\II+\u\u]h_1
    +4\gamma^{-1}[\LL-\II-\u\u]h_2,&(\en\res a)\cr
  \F(\u)&=-\gamma^{-1}\K(g_0-2g_1).&(\+b)\cr}$$
Equations (\diffeqs--\res) together with the definitions, Eqs.~(\opdef)
and~(\laplace), provide the differential formulation in the relativistic case.

In order to proceed further analytically,
it is useful to decompose the distribution function and the potentials in
spherical harmonics, e.g.,
$$f(u,\theta,\phi)=\sum_{n=0}^\infty\sum_{m=-n}^n
  f_{nm}(u)P_n^m(\cos\theta)\exp(im\phi).\eqno(\e)$$
Here $u=|\u|$ (different from the convention used earlier), $\theta$
is the polar angle, and $\phi$ is the azimuthal angle. The equation
$[L-\alpha]g=f$ is equivalent to the system of separated equations
$[L_n-\alpha]g_{nm}=f_{nm}$, where
$$[L_n-\alpha]y=(1+u^2)\frac{d^2y}{du^2}+(2u^{-1}+3u)\frac{dy}{du}
  -\biggl(\frac{n(n+1)}{u^2}+\alpha\biggr)y.\eqno(\e)$$
After the change of variable $x=\sinh^{-1}\!u$ and the change of unknown
$z=(\sinh x)^{-n}y$, then the equation $[L_n-\alpha]y=w$ transforms to
$[\DC_n-a^2]z=(\sinh x)^{-n}w$, where $a^2=\alpha+1$ and
$$[\DC_n-a^2]z=\frac{d^2z}{dx^2}+2(n+1)(\coth x)\frac{dz}{dx}
  +\bigl((n+1)^2-a^2\bigr)z.\eqno(\e)$$
The solution to the homogeneous equation $[\DC_n-a^2]z=0$ is required in
order to construct a Green's function for the problem. To obtain this solution
we note the following recurrence: If $z_{n-1,a}$ solves $[\DC_{n-1}-a^2]z=0$,
then $z_{n,a}=(\sinh x)^{-1}(d/dx)z_{n-1,a}$ solves $[\DC_n-a^2]z=0$.
Furthermore, for $n=-1$ the homogeneous equation is trivial to solve.
However, the recurrence breaks down in the case that $a$ is an integer. If
$a=n$, then $z_{n-1,a}=1$ solves $[\DC_{n-1}-a^2]z=0$, and differentiation
produces the null solution to $[\DC_n-a^2]z=0$. The recurrence must then be
restarted from the general solution to $[\DC_n-n^2]z=0$, which is
$$z_{n,n}=(\sinh x)^{-2n-1}\biggl(C_1+C_2\int_0^x(\sinh x')^{2n}dx'\biggr).$$
The integral that occurs here can be expressed in closed form.

The Green's function allows us to reduce the separated ordinary differential
equations to quadrature. An important special application for these results
is in the treatment of collisions off an equilibrium background
distribution.  Assuming that $f_b$ is a stationary Maxwellian with density
$n_b$ and temperature $T_b$ and that the energy of the colliding particles
greatly exceeds $T_b$, we obtain
$$\eqalign{D_{uu}&=\Gamma_{ab}
{K_1\over K_2} {u_{tb}^2\over v^3}
\biggl(1-{K_0\over K_1} {u_{tb}^2\over \gamma^2c^2}\biggr),\cr
D_{\theta\theta}&=\Gamma_{ab}{1\over 2 v}\biggl[1-{K_1\over K_2}
\biggl({u_{tb}^2\over u^2}+{u_{tb}^2\over \gamma^2c^2}\biggr)+{K_0\over K_2}
{u_{tb}^2\over u^2}{u_{tb}^2\over\gamma^2c^2}\biggr],\cr}$$
and $F_u=-(m_av/T_b)D_{uu}$.  (The other components of $\DD$ and $\F$
vanish.)
Here we have put the expressions for $\DD$ and $\F$ into dimensional form
as in Eqs.~(\dfeq), $K_n$ is the $n$th-order Bessel function of the second
kind, the argument for the Bessel functions is $m_bc^2/T_b$,
$u_{tb}^2=T_b/m_b$, and
$\Gamma_{ab}=n_bq_a^2q_b^2\log\Lambda_{ab}/(4\pi\epsilon_0^2m_a^2)$. The
errors are exponentially small in $u/u_{tb}$.

To conclude, we have presented a differential formulation for the Beliaev and
Budker \ref\beliaev\ relativistic collision integral. This permits the rapid
numerical evaluation of the collision term. A decomposition into spherical
harmonics is useful in carrying out analytical work. It also
provides a convenient method for calculating the boundary conditions for
the potentials.

We are grateful to N. J. Fisch for several enlightening discussions. This work
was supported by DoE contract DE--AC02--76--CHO--3073.

\bigbreak

\Ref. \landau. L. D. Landau, \jour{Phys. Z. Sowjet.}10(1936)154*.

\Ref. \beliaev. S. T. Beliaev and G. I. Budker,
\jour{Sov. Phys. Doklady}1(1956)218*.

\Ref. \rosenbluth. M. N. Rosenbluth, W. M. MacDonald, and D. L. Judd,
\jour{Phys. Rev.}107(1957)1*.

\Ref. \trubnikov. B. A. Trubnikov, \jour{Sov. Phys. JETP}7(1958)926*.

\Ref. \lifshitz. E. M. Lifshitz and L. P. Pitaevskii,
{\sl Physical Kinetics},
{\sl Course of Theoretical Physics}, Vol.\ 10
(Pergamon, Oxford, 1981), Section 50.

\Ref. \klimontovich. Yu.\ L. Klimontovich, {\sl The Statistical Theory of
Non-Equilibrium Processes in a Plasma} (Pergamon, Oxford, 1967).

\Ref. \bezzerides. B. Bezzerides and D. F. DuBois,
\jour{Ann. Phys. (NY)}70(1972)10*.

\Ref. \franz. M. R. Franz,
``The Numerical Solution of the Relativistic Bounce-Averaged Fokker-Planck
Equation for a Magnetically Confined Plasma,'' University of California Report
UCRL--96510 (April, 1987).

\bye